\documentclass[11pt]{article}
\newcommand{\Comment}[1]{{}}
\usepackage{amsfonts,amsthm,amsmath,amssymb,slashed}
\usepackage[textwidth = 430 pt, textheight = 630 pt]{geometry}

\Comment{\usepackage{color}
\definecolor{MyDarkBlue}{rgb}{0.15,0.15,0.45}
\usepackage[linktocpage=true]{hyperref}
\hypersetup{
colorlinks=true,
citecolor=MyDarkBlue,
linkcolor=MyDarkBlue,
urlcolor=MyDarkBlue,
pdfauthor={Horatiu Nastase and Amanda Weltman},
pdftitle={A natural cosmological constant from chameleons},
pdfsubject={hep-th}
}

\usepackage[numbers,sort&compress]{natbib}
\usepackage{hypernat}}
\usepackage{graphicx}
\usepackage{cite}

\newcommand\ignore[1]{}
\def\one{{\,\hbox{1\kern-.8mm l}}}

\def\a{\alpha}\def\b{\beta}

\def\lsim{\mathrel{\mathstrut\smash{\ooalign{\raise2.5pt\hbox{$<$}\cr\lower2.5pt\hbox{$\sim$}}}}}
\def\gsim{\mathrel{\mathstrut\smash{\ooalign{\raise2.5pt\hbox{$>$}\cr\lower2.5pt\hbox{$\sim$}}}}}

\newcommand{\be}{\begin{equation}}
\newcommand{\bea}{\begin{eqnarray}}

\newcommand{\ee}{\end{equation}}
\newcommand{\eea}{\end{eqnarray}}

\begin{document}

\renewcommand{\thefootnote}{\fnsymbol{footnote}}

\makeatletter
\@addtoreset{equation}{section}
\makeatother
\renewcommand{\theequation}{\thesection.\arabic{equation}}

\rightline{}
\rightline{}
   \vspace{1.8truecm}

%\begin{flushright}
% preprint nrs.
%\end{flushright}

\vspace{10pt}

%%%%%%%%%%%%%%%%%

\begin{center}
{\LARGE \bf{\sc A natural cosmological constant from chameleons}}
\end{center} 
 \vspace{1truecm}
\thispagestyle{empty} \centerline{
{\large \bf {\sc Horatiu Nastase${}^{a,}$}}\footnote{E-mail address: \Comment{\href{mailto:nastase@ift.unesp.br}}{\tt nastase@ift.unesp.br}}
{\bf{\sc and}} 
{\large \bf {\sc Amanda Weltman${}^{b,}$}}\footnote{E-mail address: \Comment{\href{mailto:amanda.weltman@uct.ac.za}}{\tt amanda.weltman@uct.ac.za}}
                                                           }

\vspace{.8cm}

\centerline{{\it ${}^a$ 
Instituto de F\'{i}sica Te\'{o}rica, UNESP-Universidade Estadual Paulista}} \centerline{{\it 
R. Dr. Bento T. Ferraz 271, Bl. II, Sao Paulo 01140-070, SP, Brazil}}

\vspace{.7cm}

\centerline{{\it ${}^b$ Astrophysics, Cosmology \& Gravity Center,}}
\centerline{{\it Department of Mathematics and Applied Mathematics, 
University of Cape Town}} 
\centerline{{\it Private Bag, Rondebosch 7700,  South Africa}}

\vspace{2truecm}

%%%%%%%%%%%%%%%%%
\thispagestyle{empty}

\centerline{\sc Abstract}

\vspace{.4truecm}

\begin{center}
\begin{minipage}[c]{380pt}
{\noindent We present a simple model where the effective cosmological constant appears from chameleon scalar fields. For a 
Kachru-Kallosh-Linde-Trivedi (KKLT)-inspired form of the 
potential and a particular chameleon coupling to the local density, patches of approximately constant scalar field potential cluster around regions of 
matter with density above a certain value, generating the effect of a cosmological constant on large scales. This construction addresses both the 
cosmological constant problem (why $\Lambda$ is so small, yet nonzero) and the coincidence problem (why $\Lambda$ is comparable to the 
matter density now).}
\end{minipage}
\end{center}

\vspace{.5cm}

\setcounter{page}{0}
\setcounter{tocdepth}{2}

\newpage

%\tableofcontents
\renewcommand{\thefootnote}{\arabic{footnote}}
\setcounter{footnote}{0}

\linespread{1.1}
\parskip 4pt

%{}~
%{}~

\section{Introduction}
\ \ \ \ \ 

The cosmological constant problem is one of the most challenging problems in theoretical physics today. Indeed, the problem is twofold. 
Firstly  the observed cosmological constant today is about 123 orders of magnitude lower than the natural value implied by quantum loop 
corrections, the uppermost cut-off for effective field 
theory, namely the Planck scale. In the days before the observation of the cosmological constant, when it was thought to be zero, the problem 
was easier, since 
some kind of yet undiscovered symmetry could perhaps force it to be zero. Supersymmetry for instance alleviates the problem a bit, since in exact global 
supersymmetry $\Lambda=0$, so broken supersymmetry requires that the cosmological constant be of the order of the supersymmetry breaking 
scale, instead of 
the Planck scale. Moreover unbroken supergravity (local supersymmetry) requires that any cosmological constant be negative.

The observation of a non-zero, positive cosmological constant dashed these hopes, and introduced yet a second more philosophical puzzle, 
the coincidence problem: why is the cosmological constant (which should be constant for all times) of the order of the matter density {\rm today}, 
when cosmologically there is nothing 
particularly special about the moment in time we happen to live in. Experimentally, $\rho_\Lambda$ is about twice the density of dark matter today. 
In Copernican terms, we appear to live at a special time in the history of the universe when $\rho_\Lambda$, $\rho_{\rm DM}$ and $\rho_{\rm matter}$ 
are all comparable. 

It is then perhaps natural to think of a dynamical scalar field which happens to now be in a region of the potential which is almost constant but has a 
very small value (though explaining that very small value is not easy), an idea known generically as quintessence. Yet such a scalar must be very 
light, and there are very strong constraints on light scalars from gravity, as these would generate an, as of yet unobserved, fifth force. Chameleon 
scalars \cite{Khoury:2003aq,Khoury:2003rn} were introduced as a way to avoid those constraints: they are scalars whose mass depends on the 
local matter density, so on Earth the fields are very massive, avoiding laboratory gravity experimental constraints, as well as those from lunar laser 
ranging etc, (see \cite{Khoury:2003rn} for further discussions). The coupling of chameleons with the local density 
takes the form
\be
V_{eff}(\phi)=V(\phi)+\rho A(\phi)\label{veff}
\ee
and we will see that rather generally the coupling function $A(\phi)$ can be written as 
\be
A(\phi)=e^{g\frac{\phi}{M_{\rm Pl}}}\label{coupmatt}
\ee
where ${M_{\rm Pl}}$ is the reduced Planck mass and $g$ is the coupling between the scalar $\phi$ and the matter type in the energy density $\rho$. 
This generic form makes contact with Brans-Dicke theories, and $g$ is defined by the small $\phi$ limit as the coupling; it is the form usually 
assumed in chameleon theories. On the other hand, on planetary and Solar System scales, the scalar force is suppressed due to the fact that the 
scalar profile only varies within a thin shell inside large bodies, hence only the mass within this 
thin shell effectively interacts via the scalar fifth force (see \cite{Khoury:2003aq,Khoury:2003rn,Hinterbichler:2010wu} for more details). 
This leaves the possibility that the chameleon scalars have some interesting behaviour on lower density scales such as vacuum 
\cite{Upadhye:2009iv,Steffen:2010ze}, space \cite{Khoury:2003rn} (within the Solar System: outside the atmosphere, yet nonzero density)
and on cosmological scales. 

Generically, quintessence is strongly constrained by both the allowed variation of the masses of fundamental particles and the low mass 
required of a quintessence field to drive dark energy \cite{Steinhardt:1999nw} (see \cite{Brax:2004qh} for an analysis within the context 
of the chameleon). Theoretically as well, one would still need to explain why the value of the potential at the minimum is so small, which is non-trivial.

In this paper we present a new approach to the cosmological constant problem, based on the chameleon scalar idea. The potential for 
the chameleon scalar that we will choose is phenomenologically motivated, based on the KKLT-inspired models used in 
\cite{Hinterbichler:2010wu,HNsoon,HNAWsoon}. 

However, unlike \cite{Hinterbichler:2010wu}, when the coupling $g$ between the chameleon and the local matter density was a fixed 
number of order one given by string theory, we will allow an arbitrary value for $g$, and find that we need a very large value for the 
coupling in our model, which is allowed in chameleon models, see \cite{Mota:2006ed}. The potential has a minimum which we set exactly at $V=0$. 

Within this model, we will find that the value for the scalar field potential is approximately constant near concentrations of matter 
(with density greater than a minimum density), and the fact that the value of this potential is so small and comparable with the density 
of matter arises simply from the condition of minimization of the effective potential (\ref{veff}).\footnote{Note that within chameleon-type 
models it was argued that the chameleon, stabilized at the ({\em average}) $\rho$-dependent minimum of $V_{eff}$, acts as a quintessence 
field, see for instance \cite{Biswas:2004be} for an early example  in supergravity. This is not what we propose here; here the dark energy 
is approximately {\em constant in time}, yet lumped around matter distributions.} In this way we translate both the issue of the smallness 
of the cosmological constant and the coincidence problem ($\rho_\Lambda\sim \rho_{DM}$) into just choosing the shape of the potential, 
which we argue is quite natural. 
Of course, we still have the old cosmological constant problem: why don't quantum corrections affect the $V=0$ value of the minimum?

\section{The model}

The universal coupling to the matter density in (\ref{veff}) appears because the metric that couples universally to matter is not the Einstein frame
metric $g_{\mu\nu}$, but the metric
\be
\tilde g_{\mu\nu}=A^2(\phi)g_{\mu\nu}.
\ee
Such a situation appears naturally in Kaluza-Klein (KK) compactifications, when the relation between the higher dimensional metric $ds_D^2$ and the lower
dimensional metric $ds_d^2$ is of the type
\bea
ds_D^2&=&R^2(\phi)ds_d^2+g_{mn}dx^mdx^n+...\cr
&\equiv& g_{MN}dx^Mdx^N\cr
ds_d^2&=&g_{\mu\nu}dx^\mu dx^\nu\cr
R&=&\Delta^{-\frac{1}{d-2}};\;\;\; \Delta=\sqrt{\det g_{mn}},
\eea
so $R$ is a modulus for the volume of compactification. If matter couples naturally to the D-dimensional metric $ds_D^2$, we obtain 
\be
A(\phi)=\frac{R(\phi)}{R_*},\label{aphi}
\ee
where $R_*$ is a particular value for $R$, close to the average value of $R$ in the Universe, to be defined shortly. The exact form of the kinetic term 
for $R$ depends on the details of the compactification, but in general it is such that it leads to (\ref{coupmatt}), therefore the canonical scalar is 
\be
\phi=\frac{M_{\rm Pl}}{g}\ln\frac{R}{R_*}.\label{canscalar}
\ee
The mechanism described here depends only on having (\ref{coupmatt}) and (\ref{aphi}), not on the fact that $R$ is obtained from KK compactification as 
the volume modulus, but we can use the KK compactification ansatz to 
motivate the form of the potential $V(R)$. Indeed the volume of the extra dimensions must be stabilized at some value, 
therefore we can approximate the potential by a quadratic 
\be
V(R)=M_{\rm Pl}^4\left[-\a (R-R_*)+\b (R-R_*)^2+\frac{\a^2}{4\b}\right],\label{quadpot}
\ee
where $\a,\b>0$ are both dimensionless, around the minimum at
\be
R_{\rm min} = R_*+\frac{\a}{2\b}\label{Rmin}.
\ee
The constant in the potential (\ref{quadpot}) was chosen such that the minimum is at $V(R_{\rm min})=0$, and the value $R_*$ was introduced 
such that (\ref{quadpot}) is valid only for $R>R_*$. For $R<R_*$, we assume that the potential is well approximated by a very steep exponential, 
\be
V(R)= M_{\rm Pl}^4\, v\, \left\{\left[e^{\gamma (R^{-k}-R_*^{-k})}-1\right]+\frac{\a^2}{4v \b}\right\}\,,
\label{steep}
\ee
with $\gamma,k>0$ and $v>0$ all dimensionless. This form is for instance the leading exponential (at small $R$) arising from the KKLT-like potential 
generated by a superpotential $W=W_0+Ae^{-ia\varrho}$, with $a<0$ instead of KKLT's $a>0$, as explained in \cite{Hinterbichler:2010wu}, so it is a rather 
natural possibility. Here $\varrho =i\sigma$ has an imaginary part $\sigma$ related to the
scalar $R$ as $\sigma=R ^{-2}\sqrt{\pi}$ for $n\leq 4$ large extra dimensions, and
$\sigma=R^{-8/n}$ for $n\geq 4$. $a<0$ can be achieved even in the context of KKLT
by adding gluino condensation on a D9-brane with magnetic flux \cite{Abe:2005rx}, as well
as being needed in a more general context because of T-duality on a gaugino
condensation potential \cite{Quevedo:1996sv}.
The parameter $v$ is included here to allow us to fix the value of
$\frac{dV}{dR(R_*)}$ for $R<R_*$ independently of $R_*$.

Minimizing the effective potential (\ref{veff}), one finds
\be
R_*\frac{dV}{dR}(R)=-\rho(R),
\ee
so for $R<R_*$, in the steep exponential side of the potential, using the fact that $R/R_*\simeq 1$, we find
\be
\frac{\rho(R)}{\rho_*}\simeq e^{\gamma(R^{-k}-R_*^{-k})}=\frac{V(\rho)+vM_{\rm Pl}^4-V_0}{vM_{\rm Pl}^4}
\ee
where we have used the implicit dependence $\rho(R)$ to denote by $V(\rho)$ the potential at the minimum value of the effective potential, and we have 
denoted by $\rho_*=\rho(R_*)$, i.e. the minimum density so that we are in the region $R\leq R_*$. We then have
\be
V(\rho)\simeq vM_{\rm Pl}^4\frac{\rho-\rho_*}{\rho_*}+V_0
\ee
where $V_0=V(R_*)=M_{\rm Pl}^4\a^2/4\b$.

At $R=R_*$, we can equate both the value of the potential, and of the derivative in (\ref{quadpot}) and in (\ref{steep}). Using the potential for $R>R_*$,
we have
\be
\frac{\rho_*}{R_*}=-\frac{dV}{dR}(R_*)= \a M_{\rm Pl}^4\label{alpharho}
\ee
leading to 
\be
V_0=M_{\rm Pl}^4\frac{\a^2}{4\b}=\frac{\rho_*}{4}\frac{\a}{\b R_*}.
\ee
Using the potential for $R<R_*$ on the other hand, we obtain
\be
\frac{\rho_*}{R_*}=-\frac{dV}{dR}(R_*)=vM_{\rm Pl}^4\gamma k R_*^{-k-1}\Rightarrow v=\frac{\rho_*}{M_{\rm Pl}^4\gamma k R_*^{-k}}\label{vrho}
\ee

We can now input experimental constraints on the parameters. In \cite{Hinterbichler:2010wu}, a constraint on $\gamma k R_*^{-k}$ was found from 
Earth laboratory experiments. The constraint was given for $g\sim {\cal O}(1)$, but we now write it for general $g$, as 
\be
\frac{R_*^k}{\gamma k}\left[\log_{10}\left(\frac{\gamma k}{R_*^k}\right)-24+\log_{10} g^2\right]\lsim g 10^{-29}.\label{labconstr}
\ee
On the other hand, from the condition that the Milky Way Galaxy be screened (it has a thin shell), 
\be
\left(\frac{3\Delta {\cal R}}{{\cal R}}\right)_{\rm G}  = \frac{\phi_{\rm cosmo} - \phi_{{\rm solar}\; {\rm system}}}{2gM_{\rm Pl}\Phi_{\rm G}} < 1\,,
\label{galaxyscreened}
\ee
where $\Phi_{\rm G}\sim 10^{-6}$ is the Newtonian potential of the galaxy, $ \phi_{\rm cosmo}$ and  $\phi_{\rm solar}$ are the values of the 
scalar field on cosmological and solar system scales respectively, related to $R$ as in (\ref{canscalar}) and ${\cal R}$ is the radius of the galaxy. 
We find that
\be
\ln\frac{R_{min}}{R_*}\lsim 2g^210^{-6}\label{galaxy}
\ee

Consider now the case $g\sim c \times 10^3$, with $c=$a few,  then (\ref{labconstr}) becomes
\be
\gamma k R_*^{-k}\gsim g^{-1}10^{30}=c^{-1}10^{27}
\ee
and (\ref{galaxy}) becomes
\be
\frac{\a}{\b R_*}\lsim 4g^2 10^{-6}\sim 4 c^2\label{abbound}
\ee

We finally get
\be
V_0\lsim c^2\rho_*;\;\;\;
v\lsim \frac{c\times 10^{-27}\rho_*}{M_{\rm Pl}^4}
\ee

leading to a potential in the $R<R_*$ -i.e., $\rho>\rho_*$- region (in the case we are close to saturating the bounds) 
\be
V\simeq b\times 10^{-27}(\rho-\rho_*)+d\; \rho_*,\label{finalpot}
\ee
with $b\sim$ a few, and $d\lsim c^2\sim$ a few, to be constrained better from experiments shortly. Note that the value of $b$ is irrelevant, all that
matters is that changing $\rho$ has almost no effect on the value of $V$ for $\rho>\rho_*$, which stays close to $d\rho_*$.
Finally, note that in order to have a consistent picture, there must be regions in the Universe which are on the quadratic piece of the potential, and 
that requires $\rho_*>\rho_\Lambda$. (\ref{alpharho}) (and (\ref{vrho}),(\ref{finalpot}) in the last inequality) then implies that 
\be
\a>\frac{\rho_\Lambda}{M_{\rm Pl}^4R_*}\sim \frac{10^{-122}}{R_*} \gsim \frac{1}{(c\gamma k)^{1/k}}10^{-122+27/k}\label{alphabound}
\ee

We now look to understand the consequences of the potential (\ref{finalpot}). Consider the case where $\rho_{\rm galaxy\;\; cluster}\sim $ a few $\rho_*$, 
(where galaxy clusters were chosen as the largest matter structure, and their density means the density of the dark and normal matter inside them), 
which means that the chameleon inside galaxy clusters is in the $R<R_*$ region, but 
a bit away from $R_*$, and moreover that 
\be
V(R\lsim R_*)\simeq V(R_*)=d\;  \rho_*\sim \rho_{\rm galaxy\;\; cluster}. 
\ee
More precisely, consider that
\be
\frac{d\;\rho_*}{\rho_{\rm galaxy\;\;cluster}}=\frac{\Omega_\Lambda}{\Omega_{matter}}\simeq \frac{72\%}{28\%}\simeq 2.5\label{expratio}
\ee
where the right hand side is the experimental value (and matter includes dark and normal matter).\footnote{Since $\rho_{\rm galaxy\;\;cluster}
\sim 10^2\rho_\Lambda$, we obtain that $d\; \rho_*\sim 2\times 10^2 \rho_\Lambda$.}
Then outside the galaxy clusters, where the density falls 
to almost zero, the chameleon drops down to $R=R_{\rm min}$, where $V(R_{min})=0$. In this way, the chameleon creates a potential 
$V$ that is almost constant 
in regions with matter (independent of the distribution of matter inside the patch), and almost zero outside. If the ratio to the matter density is taken 
as in (\ref{expratio}), this creates an extra energy contribution that accounts for the energy of the observed cosmological constant. 

Then, as the Universe expands, $\rho_{\rm galaxy\;\;
cluster}$ drops due to the Hubble expansion (the volume of the galaxy cluster expands\footnote{Of course, the region in a cluster where the density exceeds
$\rho_\Lambda$ is decoupled from the expansion, and with an even larger density it is virialized and actually contracts. But the region 
(shell) where $\rho$ drops 
to about $\rho_\Lambda$ should expand with the Hubble expansion, and inside it, the density is constant, and equal to $d\rho_*$ 
(with $d\sim c^2$). We can then choose this value of $\rho$ that expands with the Hubble expansion as being $\rho_*$. 
It is left for future work to check that $\rho_*$ is of order $\rho_\Lambda$ or a bit larger numerically. We have 
a self-consistent solution: We have a cosmological constant due to the averaging over patches of constant energy density, 
expanding with the Hubble expansion.
In turn, the patches expand with the Hubble expansion because the shell where $\rho=\rho_*$ expands due to the presence of $\rho_\Lambda$.}), 
but as long as we are still in the $R<R_*$ region, we still have 
\be
V(\rho)\simeq d\; \rho_*
\ee
therefore the value of this extra energy contribution is constant in time, i.e. it is {\em effectively a cosmological constant}. 
\footnote{Note that we have then clusters of energy density constant in space and time, which should average into a cosmological constant in the same way 
as patches of matter, increasingly localized on smaller scales (clusters of galaxies, galaxies, stars) average to the FRW solution. We hope to check this 
by numerical simulations in the future.} One can ask, is the effect of light propagation through this set of patches of cosmological constant the same 
as through an average cosmological constant, when considered at very large scales? It is a nontrivial question, but one can ask exactly the same one
in the case of the $\Lambda=0$ FRW model, since the matter distribution is also in reality not constant, but sharply peaked. The usual answer is 
that {\em on the average} we have a constant density, on exactly the same scales as for our patchwise $\Lambda$, so one can consider the average 
density. It is definitely something to check, but it is very difficult, and since it is exactly the same situation in the case of the matter density which almost
everyone takes for granted, this analysis goes beyond the scope of this paper.

We have simulated the effect of the cosmological constant with this chameleon field. Considering our original motivation, we can ask: is this 
construction {\em natural}, that is, was it natural to obtain a very small value for the observed cosmological constant, and was it natural to have a 
value for $\rho_\Lambda$ so close to $\rho_{\rm matter}$ today? The smallness of the cosmological constant is related to the 
smallness of $\rho_*=R_*\a M_{\rm Pl}^4$. 
In \cite{Hinterbichler:2010wu,HNsoon,HNAWsoon} it was shown that we can obtain a potential like the desired phenomenological 
potential from the KKLT construction
with large extra dimensions, and the required values for $R_*$ and $\a,\b,v$ are all obtained due to the $e^{ia\varrho}$ exponentials in the superpotential
and the large extra dimensions. So in that particular case, the naturalness of the small cosmological constant would be reduced to the naturalness of 
large extra dimensions. As an example, consider eq. 6.13 in \cite{Hinterbichler:2010wu}, which says that 
$\rho_*=\a R_*M_{\rm Pl}^4\sim A^2|a|^2/M_{\rm Pl}^2 e^{|a|(\sigma_*+\sigma_{\rm min})}\sim M_{\rm Pl}^4 
e^{|a|(\sigma_*+\sigma_{\rm min}-2\sigma_0)}$, with $|a|\sigma=R^{-k}$ and $A$ being the constant in the superpotential below eq. (\ref{steep}). 
We see that the smallness of $\rho_*$ (or $\a$) is reduced to the fact that the large extra dimensional volume $|a|\sigma$ appears in an 
exponent, and so its very small variations (between $\sigma_*, \sigma_{\rm min}$ and the constant $\sigma_0$) are much amplified. 
Of course, the model there was
a string theoretic model, with $g\sim {\cal O}(1)$, whereas we want $g\sim c \times 10^3$, so we can't use it directly, but we just presented it as an example
of the mechanism. The naturalness of the large extra dimensions itself is debated, but mechanisms like the warping used by Randall-Sundrum argue to make 
it natural.

As for the coincidence problem, an explanation for the fact that $\rho_\Lambda$ is close to the matter density $\rho_{\rm matter}$ today is more nuanced. 
The potential is correlated to the matter energy density, $V=V(\rho)$, but is approximately constant, at the value $d\;\rho_*$, and the fact that this is 
close to $\rho_{\rm matter}$ today is still somewhat coincidental. 

However, we also have another feature that is different from other constructions of an effective cosmological constant. Though it is was well approximated by a 
cosmological constant {\rm until now}, in the near future, this potential contribution will drop to zero. Indeed, once $\rho_{\rm galaxy\;\;cluster}$ drops 
below $\rho_*$, (\ref{finalpot}) will not be valid anymore. In fact, asymptotically, as $\rho_{\rm galaxy\;\;cluster}$ becomes very small, eventually $R$ 
will settle at $R_{\rm min}$, with $V(R_{\rm min})=0$, i.e. with no dark energy at all. So from this point of view, the coincidence is less drastic: as soon 
as $\rho_\Lambda$ becomes comparable to $\rho_{\rm matter}$ we can observe it, but it also means that we are close to the point where it will start disappearing.
After that, we will still have some dark energy, though it will take the form of a decreasing potential energy contribution. 
Note that there should be other ways to distinguish between our $V$ and a cosmological constant, in particular by focusing on the edge region of the 
patches of constant $V$, where $V$ drops to zero, so one should see a difference in the motion of matter and the propagation of light. However, this is a 
complicated issue, that could be analyzed by numerical simulations, and as such falls outside the scope of this paper. We hope to come back to it in the future.

To close the discussion of the model, we will review some experimental constraints on the model which were derived in \cite{Hinterbichler:2010wu} for the 
case $g\sim {\cal O}(1)$, to apply in our case of $g\sim 10^3$. We already discussed (\ref{labconstr}) and (\ref{galaxy}). Putting together (\ref{abbound})
and (\ref{alphabound}), we find 
\be
\sqrt{\b}\gsim\frac{10^3}{2g}\frac{\sqrt{\rho_\Lambda}}{M_{\rm Pl}^2R_*}
\ee
and then from  \cite{Hinterbichler:2010wu} the mass of the chameleon on the largest scales is (since $\rho_\Lambda=H_0^2M_{\rm Pl}^2$)
\be
m_{\rm cosmo}= \sqrt{2}g M_{\rm Pl}\sqrt{\b}R_*\gsim \frac{10^3}{\sqrt{2}}H_0,
\ee
independent of $g$. One thing which does change with respect to \cite{Hinterbichler:2010wu} is the range of the chameleon in various environments for 
$\rho>\rho_*$, 
\be
m\gsim 10^{15} g\frac{\sqrt{\rho}}{M_{\rm Pl}}
\ee
leading to 
\be
m^{-1}\lsim \frac{0.2 mm}{g\sqrt{\rho[g/cm^3]}}=\frac{0.2 \mu m}{c\sqrt{\rho[g/cm^3]}},
\ee
i.e., a factor of $g$ smaller range.

\section{Conclusions}

In this paper we have given an alternative to a simple cosmological constant based on chameleon scalars. Because of the chameleon coupling, a value for 
the chameleon potential energy as a function of the matter density was approximately constant above a certain value $\rho_*$, namely (\ref{finalpot}), but
zero for sufficiently small $\rho$ (close to vacuum). That means that for certain choices of parameters, we can have patches of approximately constant 
energy density around the largest matter structures, and zero outside them. The patches have constant energy in time, thus effectively simulating a 
cosmological constant, as long as the matter density of the largest matter structures stays above a certain value. Eventually, as matter will get diluted, 
the potential energy will start to drop, leaving no cosmological constant in the far future.

This mechanism generates a small cosmological constant despite the vacuum being at $V=0$, thus alleviating the cosmological constant problem. 
The small value
of the effective cosmological constant can be in principle obtained rather naturally, like in a large extra dimensions scenario. It would be interesting to 
see whether this scenario can be embedded in a consistent fundamental theory. We took the coupling $g$ to be $\sim $ a few $\times 10^3$, though in string 
theory $g$ is usually a fixed number of order 1.

{\bf Acknowledgements} 
%We would like to thank...
HN would like to thank the University of Cape Town for hospitality 
during the time this project was started and completed.
The work of HN is supported in part by CNPQ grant 301219/2010-9.
This material is based upon work supported financially (AW) by the National Research Foundation of South Africa.
Any opinion, findings and conclusions or recommendations expressed in this
material are those of the authors and therefore the NRF does not accept any liability in regard thereto.

\bibliography{lambdacham}
\bibliographystyle{utphys}

\end{document}